\documentclass[12pt, preprint]{aastex}

\usepackage{graphicx}
\shorttitle{CME-Streamer interaction - Type II} \shortauthors{Chen
et al.}
\usepackage{color}

\usepackage{lineno}
\usepackage{multirow}

\begin{document}

\title{A solar type II radio burst from CME-coronal ray interaction:
simultaneous radio and EUV imaging}

\author{Yao Chen\altaffilmark{1},
Guohui Du\altaffilmark{1}, Li Feng\altaffilmark{2}, Shiwei
Feng\altaffilmark{1}, Xiangliang Kong\altaffilmark{1},  Fan
Guo\altaffilmark{3}, Bing Wang\altaffilmark{1}, and Gang
Li\altaffilmark{4}}

\altaffiltext{1}{Shandong Provincial Key Laboratory of Optical
Astronomy and Solar-Terrestrial Environment, and Institute of
Space Sciences, Shandong University, Weihai 264209, China;
yaochen@sdu.edu.cn} \altaffiltext{2}{Key Laboratory of Dark Matter
and Space Astronomy, Purple Mountain Observatory, Chinese Academy
of Sciences, 210008 Nanjing, China} \altaffiltext{3}{Theoretical
Division, Los Alamos National Laboratory, Los Alamos, NM 87545,
USA} \altaffiltext{4}{Department of Space Science and CSPAR,
University of Alabama in Huntsville, Huntsville, AL 35899, USA}

\begin{abstract}
Simultaneous radio and extreme ultraviolet (EUV)/white-light
imaging data are examined for a solar type II radio burst
occurring on 2010 March 18 to deduce its source location. Using a
bow-shock model, we reconstruct the 3-dimensional EUV wave front
(presumably the type-II emitting shock) based on the imaging data
of the two STEREO spacecraft. It is then combined with the
Nan\c{c}ay radio imaging data to infer the 3-dimensional position
of the type II source. It is found that the type II source
coincides with the interface between the CME EUV wave front and a
nearby coronal ray structure, providing evidence that the type II
emission is physically related to the CME-ray interaction. This
result, consistent with those of previous studies, is based on
simultaneous radio and EUV imaging data for the first time.
\end{abstract}


\section{Introduction}

It has long been suspected that the interaction of a coronal mass
ejection (CME) with nearby dense coronal structures like streamers
or rays is important to solar type II radio bursts
\citep[e.g.,][]{Wild1963, Classen2002, Reiner2003, Mancuso2004,
Cho2005}. These coronal structures are featured by higher plasma
density and lower bulk flow speed (in the magnetically-closed part
of streamers) than that of the surrounding solar wind plasmas
\citep[e.g.,][]{Habbal1997, Strachan2002}. Thus when a CME
interacts with a nearby streamer or coronal ray, the associated
disturbance may propagate into a region with much lower
characteristic speed. This favors the formation or enhancement of
coronal shock, as well as the consequent electron acceleration and
type II excitation.

Previous studies along this line of reasoning were largely based
on a combined analysis of coronagraph images and radio
spectrograph data. For example, \citet{Reiner2003} suggested that
the type IIs of their study originated from dense region of the
corona, likely from streamers. \citet{Cho2007} determined the
shock heights by combining the type II spectral data from Green
Bank Solar Radio Burst Spectrometer (GBSRBS; \citet{White2006})
and the polarization Brightness (pB) data from the Mauna Loa Solar
Observatory (MLSO) MK4 coronameter \citep{Elmore2003}, and
compared them to the MK4 CME-front heights. They concluded that
the type II burst was generated at the interface of the CME flank
and the streamer. In a follow-up study, \citet{Cho2008} expanded
the study to 19 events and found that nearly half of them were
probably associated with CME-streamer interaction.

In a series of studies, \citet{Feng2012, Feng2013} and
\citet{Kong2012} proposed a novel method to diagnose the type II
sources by relating specific morphological features (e.g., bumps
or breaks) of the dynamic spectra to imaging features (e.g.,
CME-driven shock propagating across a dense streamer). These
studies confirmed that the CME-shock interaction with dense
coronal structures like streamers is important to type II bursts.
However, none of the studies was based on radioheliograph imaging
data which provide the most direct evidence on the radio source
location. This is due to the scarcity of simultaneous imaging in
radio and white light/EUV wavebands. Nevertheless, existing
studies have demonstrated the importance of combining these
imaging observation as well as the spectral data in revealing the
origin of metric type II bursts and their relationship with
CME-driven shocks \citep[e.g.,][]{Bastian2001, Maia2000,
Dauphin2006, Bain2012, Zimovets2012, Carley2013}. Along this line
of observational endeavors, in this study we show a type II event
with imaging data from both the Nan\c{c}ay radioheliograph (NRH;
\citet{Kerdraon1997}) and instruments onboard the twin Solar
TErrestrial RElations Observatory (STEREO; \citet{Kaiser2008}) and
the Solar and Heliospheric Observatory (SOHO; \citet{Domigno1995})
spacecraft. This allows us to pinpoint the type II source location
that is found to correspond to a CME-coronal ray interface.

\section{Spectral and imaging observations of the type II burst on 2010 March 18}\label{sec2}

The type II burst was imaged by NRH at 173.2 and 150.9 MHz, also
recorded by the Humain spectrograph from 45 - 80 MHz and 163 - 387
MHz and by OOTY (http://soleil.i4ds.ch /solarradio/) from 80 - 163
MHz, with a temporal resolution of 0.25 s. The spectral
resolutions vary irregularly with frequencies. For the data below
80 MHz, the resolution varies from $\sim$0.06 to 1 MHz, and the
higher frequency data have a resolution from 1 to 3 MHz. A
combination of these data is shown in Figure 1. The contours in
the NRH images represent the 80\% (blue), 90\% (orange), and 95\%
(white) of the corresponding intensity maximum.

The type II burst lasted for about 1.5 min from 11:19:30 to
11:21:00 UT. The start (end) frequency of the F branch is $\sim$80
(60) MHz on the lower band. This burst shows band-splitting
feature on both the fundamental (F) and harmonic (H) branches
\citep{Smerd1974, Nelson1985}. The splitting width, i.e., the mean
frequency ratio of the upper and lower bands, is estimated to be
$\sim$1.17, and the average drift (F band) is $\sim$0.2 MHz
s$^{-1}$. Both are typical values in this frequency range
\citep[e.g.,][]{Vrsnak2001}.

The radio source centroid location at the two NRH frequencies can
be told from Figures 1(a) and (b). Its uncertainty can be
estimated with the 95\% contour of the NRH flux intensity, as
usually done in literature \citep[see, e.g.,][]{Bain2012,
Zimovets2012}. According to the 150.9 MHz data, the position of
the type II source is in a range of (y = -0.48$\pm$ 0.02, z=0.26
$\pm$ 0.02) in units of solar radii (R$_\odot$), expressed in the
Heliocentric Earth Ecliptic (HEE) coordinate system with the
x-axis pointing from the solar center towards the Earth, z
perpendicular to the Earth orbital plane (positive north), and y
completing the right-handed orthogonal triad. From panel (c), the
radio flux at 173.2 MHz (blue) increased rapidly to a brightness
temperature (T$_B$) of 10$^{9}$ K within seconds. From the
accompanying animation, at 150.9 MHz and 11:18:30 UT there
appeared a bright emission with $T_B$ = 10$^{8}$ K in the same
position, indicating that the burst may start as early as 11:18:30
UT. However, no corresponding spectral feature was found due to
the lower sensitivity of spectrographs than NRH.

The type II sources observed by NRH at 150.9 MHz (11:20:00 UT) and
173.2 MHz (11:19:38 UT) were close to each other since the
uncertainty of the radio source centroid measurement ($\sim0.02$
R$_\odot$) is considerably larger than the projected moving
distance of the radio source within 20 seconds (see next section
for a wave speed measurement). It is therefore not possible to
deduce the shock speed using available NRH data.

\section{EUV/white-light data of the solar eruption}\label{sec3}

As observed by Extreme Ultraviolet Imaging Telescope \citep[EIT;
][]{Delaboudiniere1995}/SOHO and the Extreme Ultraviolet
Imager\citep[EUVI; ][]{Wuelser2004} /STEREO, the type II burst was
associated with a filament eruption, a CME, and a weak B5 X-ray
flare from a small active region (AR) at N23E22, which is
northwest of the NOAA AR11056 (N18E20). The flare start, peaking
and end times were $\sim$11:09, 11:15, and 11:18 UT, respectively,
according to the online GOES-SXR data
$\url{http://lmsal.com/solarsoft/last\_events\_20100322\_0423/gev\_20100318\_1109.html}$).

In Figure 2(a), we show the overall configuration of the
spacecraft. STEREO-A (SA) was 66$^\circ$ ahead and STEREO-B (SB)
71$^\circ$ behind the earth. The CME source was at N23E22 in the
SOHO field of view (FOV), while in the SB (SA) FOV it was at
N27W49 (N28E88). Therefore, the CME was basically observed by SA
as a limb event. In Figure 2(b) we show the EIT difference image
and the NRH intensity contours (taken from Figure 1(a)). It can be
seen that the NRH radio source lagged behind the southern-eastern
part of the EIT-EUV front, mainly due to the projection effect and
the fact that the EIT data were obtained 4 min later. In the
following section, the 3-d reconstruction of the EUV wave fronts
will be extrapolated to the times of the EIT and NRH data for a
direct comparison. The type II source seems to be close to the SA
solar limb (the red dashed line in Figure 2(b)), as will be
confirmed by further analysis.

From the EUVI 195 \,\AA{} RDIs shown in Figure 2 and the online
animation, the CME EUV front swept through a nearby ray along the
southern part. The front appeared at a time as early as 11:05:30
UT, propagated to a distance of $\sim$1.14 R$_\odot$ from the
solar center at a time (11:10:51 UT) close to the flare onset, and
reached $\sim$1.25 R$_\odot$ at 11:13:00 UT near the flare peak.
In addition, from the EUVI-A 304 \,\AA{} data shown in Figure
3(a), an associated filament eruption was observed at 11:06:15 UT.
These indicate that the EUV wave was driven by the CME, rather
than by the flare.

The ray that was swept by the CME front is relatively bright and
thus dense (see the yellow arrows in Figure 2). It is connected to
the bright loop complex as seen from SB and to a cusp-like
structure from SA. These features are similar to those of typical
streamers observed by coronagraphs, yet with a low cusp. Upon the
CME impact, the ray was strongly deflected, its brightness
increased and width decreased consequently.

Further details of the interaction can be seen from the running
difference images (RDIs) shown in Figure 3(b-i) and the
accompanying animation. The bright front along the southern part
of the eruption can be clearly recognized. We delineated four of
these fronts using blue solid circles, which is of further usage
in the Appendix for shock reconstruction. The average propagation
speed of the SA EUV front is $\sim$465 km s$^{-1}$, on which the
projection effect is not serious since the event is a limb one in
the SA FOV.

Consistent with Figure 2, the expanding front swept through the
upper part of the ray in the 5-min interval from 11:15 to 11:20
UT. The deflected ray, as well as the interaction interface, can
be clearly seen by SA. As early as 11:18 UT, the CME front may
have already touched the lower northern part of the ray. The
deflection is clearly manifested as a bright structure enclosed by
the CME front in RDIs of Figures 3(d) to 3(f). This suggests that
the expanding CME has passed by the lower part of the ray at the
corresponding moment, indicating that the CME front is not the
ejecta itself. Rather, it has the nature of a fast magnetosonic
wave or a fast shock that can propagate across magnetic
structures. At the time after panel (i), the CME front became
diffusive and hardly recognizable.

We emphasize that the EUVI images were recorded at 11:20:30 UT
from SA and 11:20:51 UT from SB. At both times, the NRH images of
the type II burst at 150.9 MHz were also available. These
simultaneous imaging at the EUV and radio wavelengths from
different vantage points are critical to our study, as shown in
the following section.

\section{Origin of the type II radio burst}\label{sec4}

Although the EUV front presents a fast-mode nature, its relation
with the type-II emitting shock is a question of intensive studies
(e.g., Biesecker et al. 2002; Klassen et al. 2000; Pohjolainen et
al. 2001; Khan {\&} Aurass, 2002; Vrsnak et al. 2006). These
studies provide important insights into the physical relation
between the type-II emitting shock and the EUV front. Based on
these studies, the demonstrated wave nature of the front, the
observed deflection of the coronal ray, as well as the fact that
the EUV front outlines the outermost envelope of the CME, we
presume that the EUV front corresponds to the type-II-emitting
shock (see also Ontiveros {\&} Vourlidas, 2009). With this
assumption, we reconstruct the 3-d surface of the EUV fronts based
on the nearly-simultaneous data of the two STEREO spacecraft. The
details of the method are presented in the Appendix. The
reconstructions are used to establish the physical relationship
between the type II burst and the CME interaction with the nearby
ray structure.

Given the successful reconstruction of the 3-d shock surface (see
the Appendix) and the knowledge of the NRH type II source, it is
straightforward to determine the 3-d coordinates of the type II
source. In Section 2, we obtained that the NRH radio source is
located at y = -0.48 $\pm$ 0.02 R$_\odot$, z = 0.26 $\pm$ 0.02
R$_\odot$ at 11:20:30 UT, the corresponding x coordinate on the
3-d shock surface can then be deduced to be x = 1.20 $\pm$ 0.03
R$_\odot$. Please see the Appendix for an estimate of the error
contributed by the reconstruction.

SA is 66$^\circ$ to the west of the Earth, after a rotation we get
the type II source location viewed in its FOV: $x_A$ = 0.04 $\pm$
0.03, $y_A$ = -1.29 $\pm$ 0.03, and $z_A$ = 0.26 $\pm$ 0.02
(R$_\odot$). The errors consist contributions from both NRH
imaging data and the reconstruction (see the Appendix for
details). Since $y_A$ and $z_A$ are both significantly larger than
$x_A$, the type II source was very close to the solar limb of SA,
as expected earlier. In Figure 4(a), this source is plotted on the
simultaneous EUVI-A image. The obtained profile of the 3-d shock
is also shown as yellow curves.

It is found that the type II source lay right on top of the
deflected bright streamer structure, i.e., basically coincide with
the CME interface with the ray. The collocation of the type II
source and the CME-streamer interface is a strong evidence
supporting that the type II is physically related to the CME-ray
interacting process.

To make a direct comparison between the NRH and EIT data which
were obtained at $\sim$11:20:00 and 11:24:10 UT, respectively, we
extrapolate the reconstruction result at 11:20:30 UT to the above
times. These extrapolated bow shock profiles are superposed onto
the corresponding NRH and EIT images in Figures 4 (b) and (c). We
see that the type II source is located at the southern part of the
shock front, the radio source centroid is $\sim$ 100 arcsecs away
from the shock nose, and the reconstructed shock front using the
STEREO data basically envelopes the visible part of the EIT wave
front.

\section{Conclusions and discussion}\label{sec5}

In this paper we show evidence supporting that the type II solar
radio burst occurring on 2010 March 18 was originated from the
interaction between a CME and a nearby dense coronal ray
structure. This is based on simultaneous radio and EUV imaging
data recorded at different vantage points. The NRH data were used
to constrain the projected type II source location in the Earth
FOV, and the multi-vantage point EUV data were used to reconstruct
the 3-d profile of the EUV front (presumably the type-II emitting
shock). We find that the type II source lay closely to the CME-ray
intersection region. This study demonstrates the importance of CME
interaction with dense coronal structures (like rays or streamers)
to metric type IIs.

The suspect that CME-streamer interactions may be important to
metric type IIs \citep[e.g.][]{Wild1963, Classen2002, Reiner2003,
Mancuso2004, Cho2005} stems from following considerations.
Firstly, the CME-related disturbance may frequently run into a
nearby density structure along the CME flank due to its lateral
expansion. This interaction region likely corresponds to a
quasi-perpendicular shock geometry, if a shock exists. Shocks of
such geometry have been predicted to account for efficient
electron acceleration \citep{Holman1983, Wu1984}. Secondly, it is
expected that when a CME disturbance propagates into a dense
coronal structure with low Alfv\'enic speed, the disturbance may
steepen into a shock, or a preexisting shock may get enhanced in
strength, also favoring electron acceleration. Thirdly, it is
suggested that the closed field topology within the streamer, when
swept by an outward-propagating shock, may give rise to a
collapsing magnetic trap configuration \citep{Baker1987,
Zlober1993, Somov1997, Chen2013}, within which electrons are
trapped by the upstream field lines, and may return to the shock
and be accelerated repeatedly. This also favors the generation of
energetic electrons. While we do not consider the exact electron
acceleration process in this word, our study is consistent with
the above scenarios.

\acknowledgements

We are grateful to the STEREO, SOHO, NRH, and the Humain/OOTY
teams for making their data available to us. We thank the referee
for constructive comments. This work was supported by grants
NSBRSF 2012CB825601, NNSFC 41274175 and 41331068. Li Feng's work
was supported by grants NNSFC 11003047, 11233008, and BK2012889.
Gang Li's work at UAHuntsivlle was supported by NSF grants
ATM-0847719 and AGS1135432.

\appendix

\section{A bow reconstruction of the 3-d shock surface: method and result}
We assume the CME shock can be approximated by a symmetrical 3-d
bow-shock geometry (see \citet{Ontiveros2009}). The bow shock is
given by (\citet{Smith2003}, see also bowshock{\_}cloud.pro in
Solar Software),

$$
Z=h-\frac{d}{s}\times(\frac{\sqrt{X^{2}+Y^{2}}}{d})^{s}
$$
where $h$ determines the shock apex height from the CME start
point (S for short), $s$ controls the bluntness of the shock
surface, and $d$ describes how wide the shock shape is. The
equation is given in a local Cartesian coordinate system with S
taken to be the origin, the bow symmetrical axis is along Z
pointing from S to the bow apex (A), and X and Y define the plane
perpendicular to Z. The above parameters define the shape and
height (measured from S) of the bow. The bow axis, when measured
in the Earth Ecliptic spherical coordinate system, is further
constrained by other two pairs of angular parameters. The first
pair gives the co-latitude and longitude $(\theta,\psi)$ of S, and
the other pair of co-latitude and longitude
$(\theta^\prime,\psi^\prime)$ gives its pointing direction.

We first measure the shock front locations from EUVI images. More
measurements on the SA data are used since in SA the CME is more
clear and at the limb. The best-fit solution was deduced by
minimizing the standard deviation of the measurements and the
reconstructed shock front locations. This is done for the EUV data
at 11:20:30 and 11:25:30 UT for SA and 11:20:51 and 11:25:51 UT
for SB. The influence of the SA-SB time difference (21 s) is
reduced by extrapolating the SB measurements slightly backwards by
a distance of 1.3 $\times$ 10$^{-2}$ R$_{\odot}$, which is given
by measuring the average EUV-front speed in SB FOV ($\sim$450 km
s$^{-1}$). The extrapolated measurements are used for
reconstructions. The bows thus obtained are superposed onto the
original EUV images in Figure 5, where the bows for SB have been
extrapolated outwards to account for the 21 s time difference. We
see that the above technique provides a nice fitting to the EUV
fronts observed by both STEREO spacecraft.

To estimate the associated error on the radio source location, we
first conduct reconstructions for $>$10 independent measurements
of the EUV wave fronts, and deduce a distribution of the x
coordinate of the type-II source. The resulting 1-$\sigma$ spread
of x is $\sim$0.01 R$_\odot$, taken to be the error of the
reconstruction. Taking this into account, the total error of the x
coordinate becomes $\pm$ 0.03 R$_\odot$ in the NRH FOV, and that
of the x and y coordinates become $\pm$ 0.03 R$_\odot$ in the SA
FOV.

\begin{figure}
\includegraphics[width=0.95\textwidth]{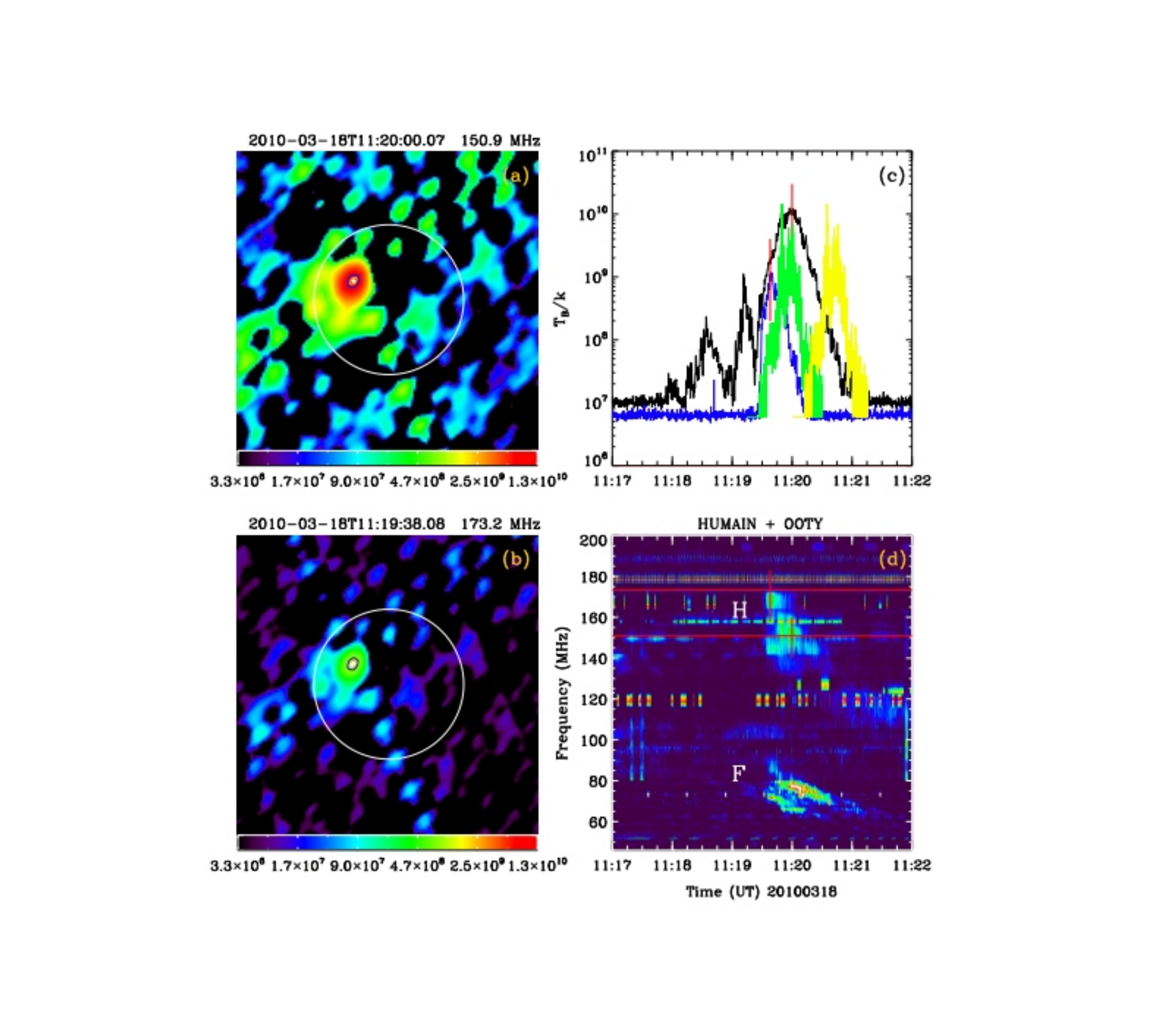}
\caption{Radio data for the event. (a, b) NRH images at 150.9 and
173.2 MHz; (c) Temporal evolution of T$_B$ at 173.2 MHz (blue) and
150.9 MHz (black). (d) Dynamic spectra recorded by Humain for
45-80 MHz and 163-387 MHz and OOTY for 80-163 MHz. The vertical
lines in (c) and (d) indicate the times of the two radio images
shown in (a) and (b). The horizontal lines in (d) represent the
two NRH frequencies. Note the OOTY spectrum has been shifted by 45
s to keep the spectral data consistent. This is determined by
comparing the original (yellow) and temporally-shifted (by 45 s,
green) OOTY flux profiles at 150.9 MHz (see panel (c), in
arbitrary unit) to that observed by NRH at the same frequency. (An
animation and a color version of this figure are available
online.)}\label{Fig1}
\end{figure}

\begin{figure}
\includegraphics[width=0.95\textwidth]{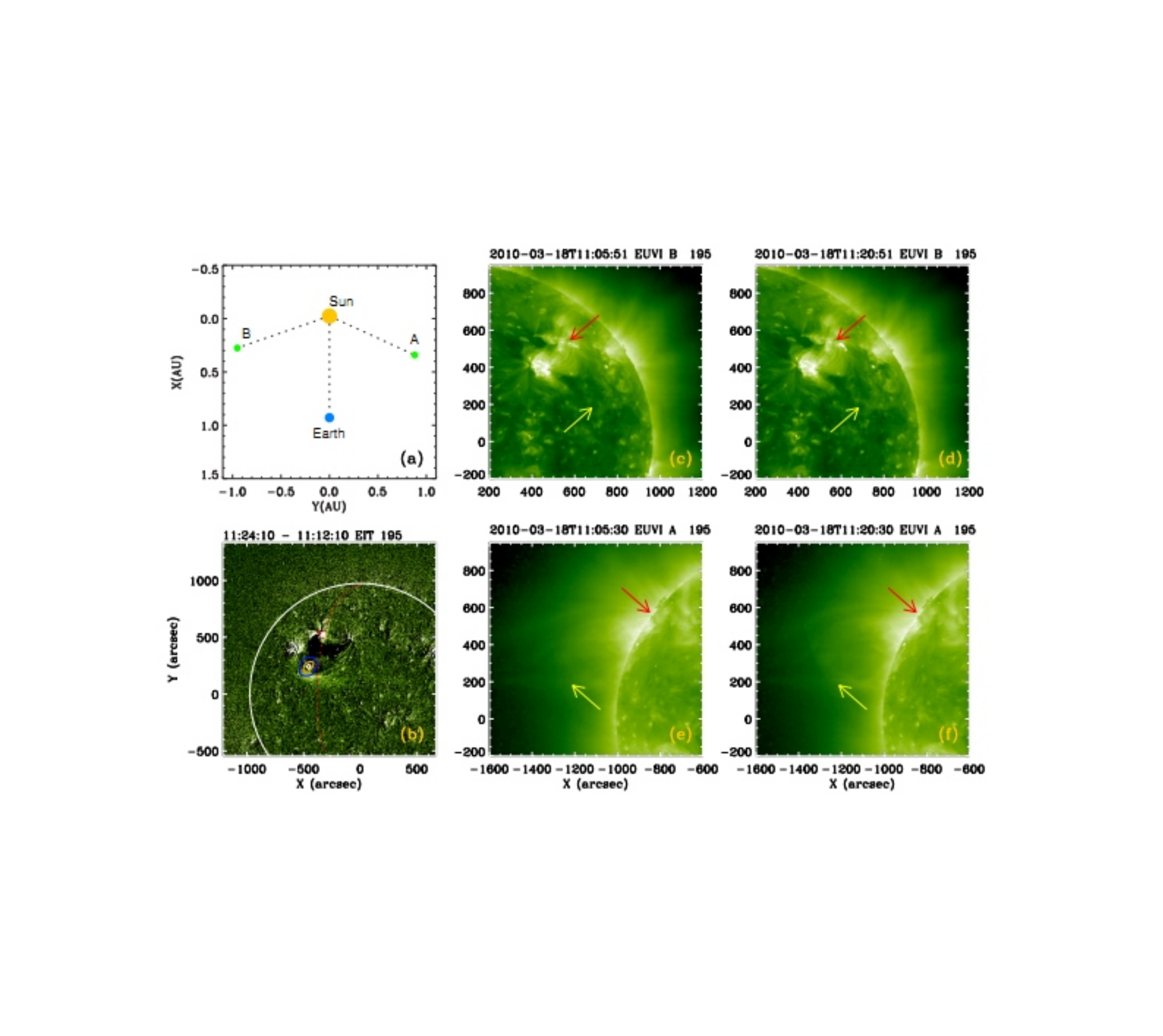}
\caption{(a) The overall configuration of SOHO, SA, and SB. (b)
The EIT difference image. (c-f) The 195 \,\AA{} images obtained by
EUVI -A and -B. The yellow arrow points to the deflected ray
structure and red to the eruption source. In panel (b), the NRH
contours shown in Figure 1(a) are plotted, the red solid circle
represents the CME source. The red dashed line presents the SA
solar limb. (An animation for this figure and Figure 3 and a color
version of this figure is available online.)}\label{Fig2}
\end{figure}

\begin{figure}
\includegraphics[width=0.95\textwidth]{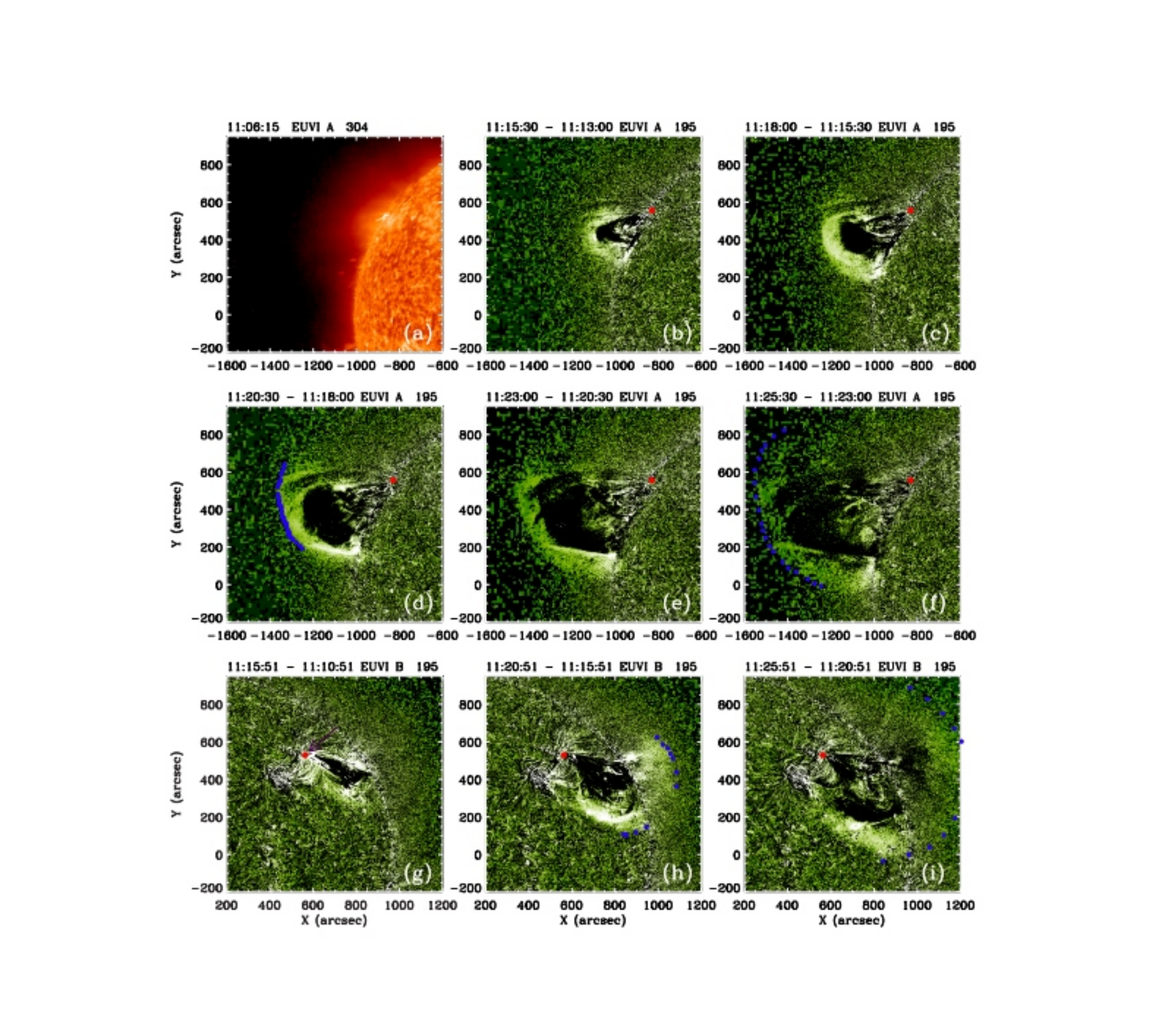}
\caption{The 304 \,\AA{} direct image of EUVI-A (a) and 195
\,\AA{} RDIs of EUVI-A (b-f) and -B (g-i). The blue solid circles
represent our measurements of the EUV front. The red solid circle
denotes the CME source. (A color version of this figure is
available online, an accompanying animation has been presented
along with Figure 2.)} \label{Fig3}
\end{figure}

\begin{figure}
\includegraphics[width=0.95\textwidth]{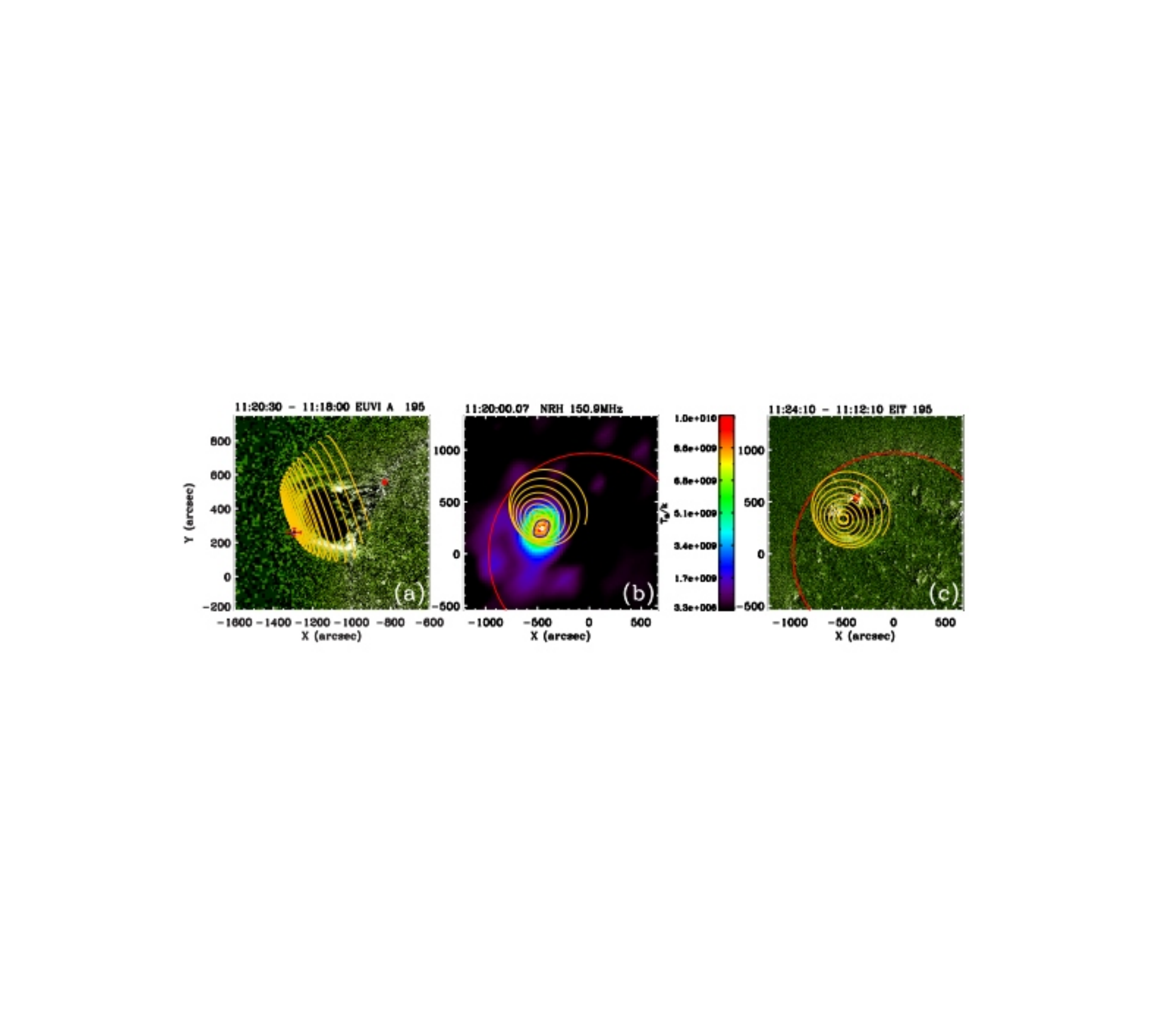}
\caption{The SA EUVI RDI (a), the NRH and EIT direct images (b-c).
The red dot denotes the CME source, the blue dot with red bars
denotes the deduced type II source position. The yellow curves in
(a) are given by the 3-d reconstruction of the shock surface,
which are extrapolated to 11:20:00 UT in (b) and to 11:24:10 UT in
(c). (A color version of this figure is available online.)}
\label{Fig4}
\end{figure}

\begin{figure}
\includegraphics[width=0.95\textwidth]{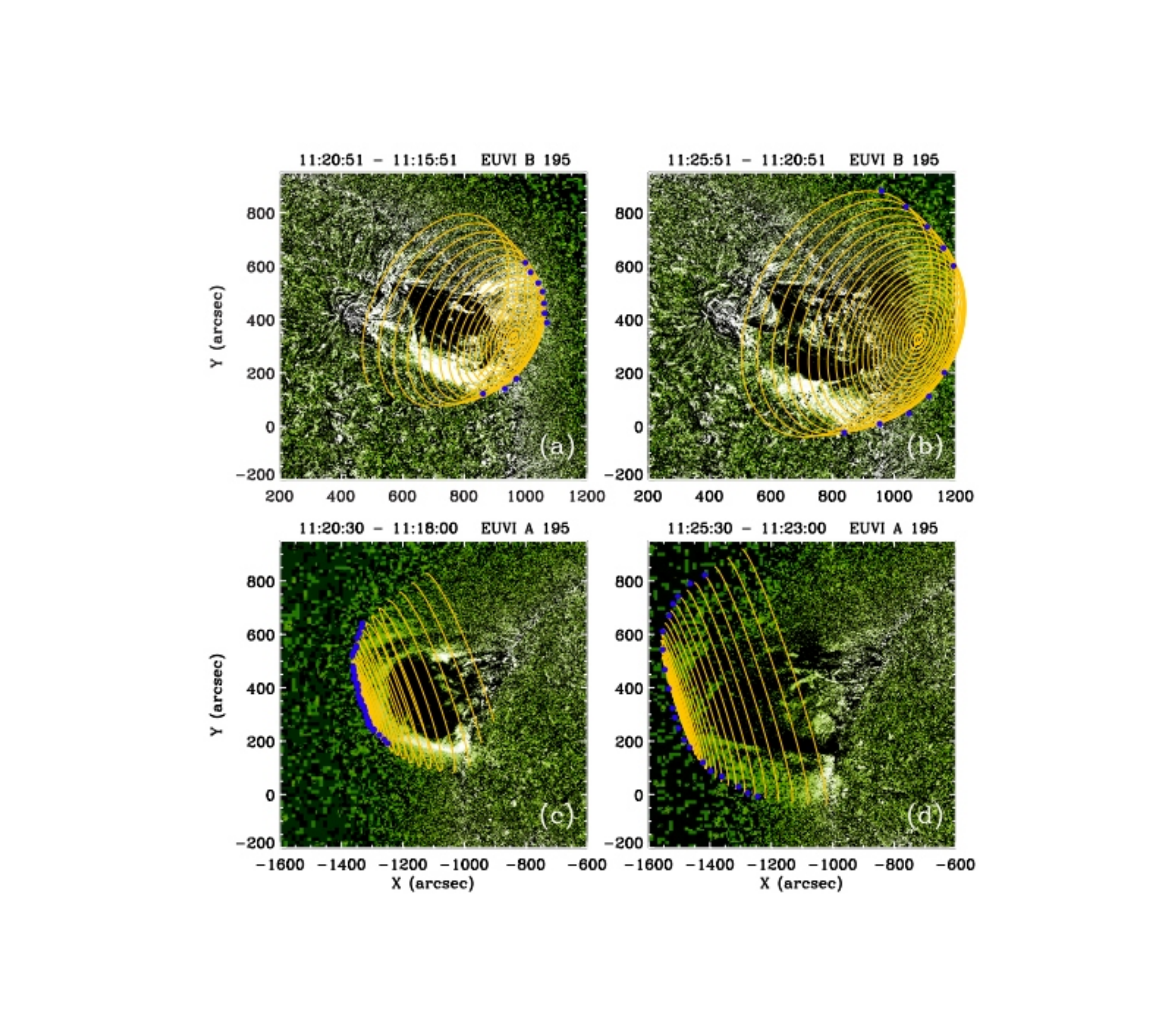}
\caption{The reconstructed shock surface (yellow curves),
superposed onto the corresponding EUVI-A and -B images. The blue
solid circles represent the measurements of the EUV front. (A
color version of this figure is available online.)} \label{Fig5}
\end{figure}
\end{document}